\documentclass[journal=jpclcd,manuscript=letter,layout=traditional]{achemso}

\usepackage{hyperref}

\usepackage{graphicx}

\usepackage{amssymb}
\usepackage{amsmath}
\usepackage{upgreek}
\usepackage{lineno}
\usepackage{gensymb}
\usepackage[utf8]{inputenc}
\usepackage{chemformula} 
\usepackage[normalem]{ulem}


\author{Simon Hettler}
\email{hettler@unizar.es}
\author{David Sebastian}
\author{Mario Peláez-Fernández}
\affiliation[LMA,UniZar]{Laboratorio de Microscopías Avanzadas, INMA, Universidad de Zaragoza, Zaragoza, Spain}
\author{Ana M. Benito}
\author{Wolfgang K. Maser}
\affiliation[ICB,UniZar]{Instituto de Carboquimica (ICB-CSIC), Zaragoza, Spain}
\author{Raúl Arenal}
\email{arenal@unizar.es}
\affiliation[LMA,UniZar]{Laboratorio de Microscopías Avanzadas, INMA, Universidad de Zaragoza, Zaragoza, Spain}
\alsoaffiliation[ARAID]{ARAID Foundation, Zaragoza, Spain}
\alsoaffiliation[INMA]{Instituto de Nanociencias y Materiales Aragón (INMA), CSIC-U. Zaragoza, Zaragoza, Spain}

\title{Evolution of graphene oxide properties during its reduction by Joule heating in in-situ transmission electron microscopy}

\begin{document}

\begin{tocentry}
\includegraphics[width=5.08cm]{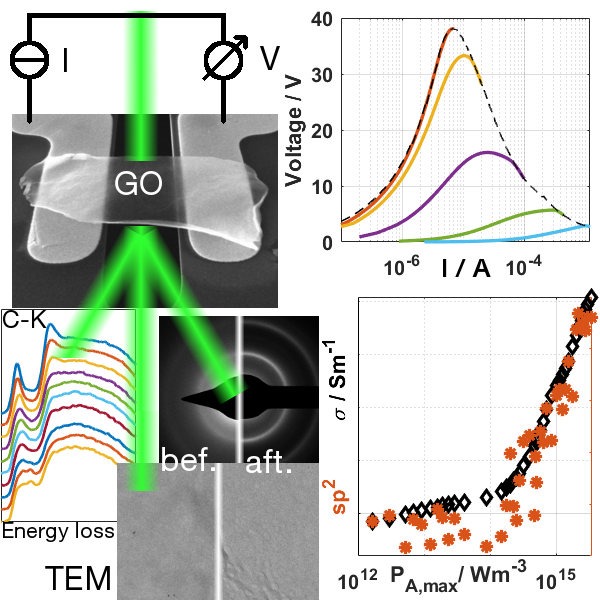}
\end{tocentry}
\begin{abstract}

Graphene oxide (GO) is reduced by Joule heating using in-situ transmission electron microscopy (TEM). The approach allows the simultaneous study of GO conductivity by electrical measurements and of its composition and structural properties throughout the reduction process by TEM, electron diffraction and electron energy-loss spectroscopy. The small changes of GO properties observed at low applied electric currents are attributed to the promotion of diffusion processes. The actual reduction process starts from an applied power density of about 2$\cdot$10\textsuperscript{14}~Wm\textsuperscript{-3} and occurs in a highly uniform and localized manner. The conductivity increases more than 4 orders of magnitude reaching a value of 3$\cdot$10\textsuperscript{3} Sm\textsuperscript{-1} with a final O content of less than 1\%. We discuss differences between the reduction by thermal annealing and Joule heating.

\end{abstract}

\maketitle

Graphene oxide (GO) strongly attracts research interest due to its use as precursor material for graphene-based material and its high flexibility in terms of functional modification offering promising applications in numerous fields \cite{Review.2012}. For a successful and tailored application, it is essential to understand the properties of GO and reduced (r)GO, which are mainly defined by the amount and types of oxygen functional groups (OFGs) and their spatial distribution within the material. Therefore, numerous studies employing different characterization techniques such as atomic force microscopy (AFM), electron microscopy (EM) or X-ray photoelectron spectroscopy (XPS) have been carried out at different states of reduction \cite{EELSGO.2016,XPSGO_rGO.2019,Liu.2018.ModelAFM}. 

The reduction process of GO may be performed chemically and/or thermally with the method of choice strongly depending on the desired rGO properties and the application under consideration. GO is essentially an electrical insulator, its conductivity increases with the state of reduction and can be described by variable-range hopping (VRH) of electrons between conductive sp\textsuperscript{2}-hybridized islands \cite{GomezNavarro.2007,JungElecCondrGO.2008}. For reaching a high degree of reduction and thus graphene-like properties, temperatures above 1000\degree C are necessary in thermal reduction processes carried out by thermal annealing in vacuum using an external heat source \cite{AcikThermalReductionIR.2011,GangulyThermalRedXPS.2011}. Instead of thermal annealing, the reduction process can as well be performed by application of Joule heating, that is by dissipation of energy in the sample using an electrical current, opening up a way to simultaneously measure its conductivity. The effect of Joule heating on the electrical characterization of C nanotubes and on the contact between metal and graphene has been studied \cite{JouleHeatingContactDong.2007,Wei.2016,Nilsson.2017} and Joule heating of conductive paths is frequently employed as external heat source in in-situ heating transmission (T)EM. Joule heating of graphene and related materials has been studied by applying voltages or currents in in-situ TEM allowing the observation of graphene sublimation \cite{Huang.2009}, the catalyst-free transformation of amorphous C to graphene \cite{Barreiro.2013} or the formation of graphene from a C-Cu needle \cite{Rosmi.2016}. Two studies showed the successful reduction of individual flakes of GO by Joule heating via in-situ TEM \cite{Xu.2011,Martin.2018} but important questions on the microscopic processes involved in Joule heating and the details of GO structural evolution remain open. Here, we present the controlled in-situ reduction of a GO film by Joule heating performed in an in-situ TEM sample holder dedicated for electrical measurements allowing the simultaneous determination of its conductivity and its structural evolution. 

The in-situ reduction and electrical measurements were carried out in a Thermo Fisher Scientific Titan\textsuperscript{3} transmission electron microscope with an aberration corrector for the objective lens, using a DENSsolutions Wildfire sample holder and a 'through-hole chip' consisting of 4 contacts located around a small hole on a membrane. The TEM studies were conducted at an electron energy of 80~keV. GO was obtained by overnight oxidation of pristine graphite flakes using a modified Hummers method as reported elsewhere. \cite{GO_SynthVictorRoman.2020}. Briefly, oxidation with KMnO\textsubscript{4}, H\textsubscript{2}SO\textsubscript{4} and NaNO\textsubscript{3} followed by addition of H\textsubscript{2}O\textsubscript{2}/H\textsubscript{2}O and subsequent washing with HCl and H\textsubscript{2}O and filtering results in a solid product (graphite oxide), which was dried at room temperature. Mild sonication (45 kHz) of the obtained graphite oxide in water during 30 minutes yields the corresponding GO aqueous dispersion in a concentration of 1~mg/mL. For sample preparation we first prepared a conventional TEM sample by drop-casting 2~$\upmu$l of the GO dispersion on a plain 1000 mesh Cu grid (hole size approximately 18x18 $\upmu$m\textsuperscript{2}) made hydrophilic in a glow discharge process. After drying at room temperature, the individual GO flakes in the dispersion pile up on the grid forming a continuous film. A Thermo Fisher Scientific Helios 650 focused-ion beam (FIB) instrument was then used to cut and transfer a piece of the GO film to the DENSsolutions chip assisted by micro-needles (Omniprobe and Kleindiek Nanotechnik) and a gas injection system (GIS). We employed several approaches and the best result was obtained by the application of two micro-needles used for GO transport and electrostatic discharging of the metal contacts, combined with focused-electron and focused-ion beam induced deposition (FEBID/FIBID) of C (precursor gas C\textsubscript{10}H\textsubscript{6}) for electrical contacting and fixation of the film to the chip. Details on FIB sample preparation may be found in the supplementary information (Fig.~S1). Figure~\ref{F:Elec}a shows a scanning electron microscopy image of the the GO film mounted on the chip.

The in-situ reduction experiment is composed of 47 individual measurements of the film with an effective area between the contacts of 6~$\upmu$m x 5~$\upmu$m. Each measurement consists of the application of a linear current sweep and simultaneous measurement of the necessary voltage using a Keithley SourceMeter 2450 (Tektronix, Keithley Instruments) from 0 up to a maximum current $I_{max}$, with $I_{max}$ increasing with the measurement number reaching a maximum value of 1.2~mA at the end of the experiment series. Table~S2 shows a list of $I_{max}$ together with other experimental parameters and obtained results of the film evolution. Each current sweep is followed by a TEM, EELS and electron diffraction (ED) analysis of the film with no electrical field applied using identical conditions and selecting a circular area with a diameter of 490~nm for ED and EELS data. The irradiation dose rate in this setup could be kept below 1~e\textsuperscript{-}s\textsuperscript{-1}\AA \textsuperscript{-2}. EELS data was acquired with a Gatan image filter (GIF) Tridiem operated at a dispersion of 0.1 eV/px, a collection angle of 19.7 mrad and a convergence angle of $<$0.1 mrad, thus under parallel illumination. Spectra of the low-loss (LL) region containing the zero-loss (ZL) peak (exposure time 10\textsuperscript{-6} s), the Si-L (0.1 s), the C-K (0.3 s) and the O-K edge (0.5 s) were automatically acquired using a custom Digital Micrograph script applying binned-gain averaging of 30 spectra \cite{BosmanBinnedGain.2008}. 

\begin{figure}[ht]
\centering
\includegraphics[width=0.67\linewidth,keepaspectratio]{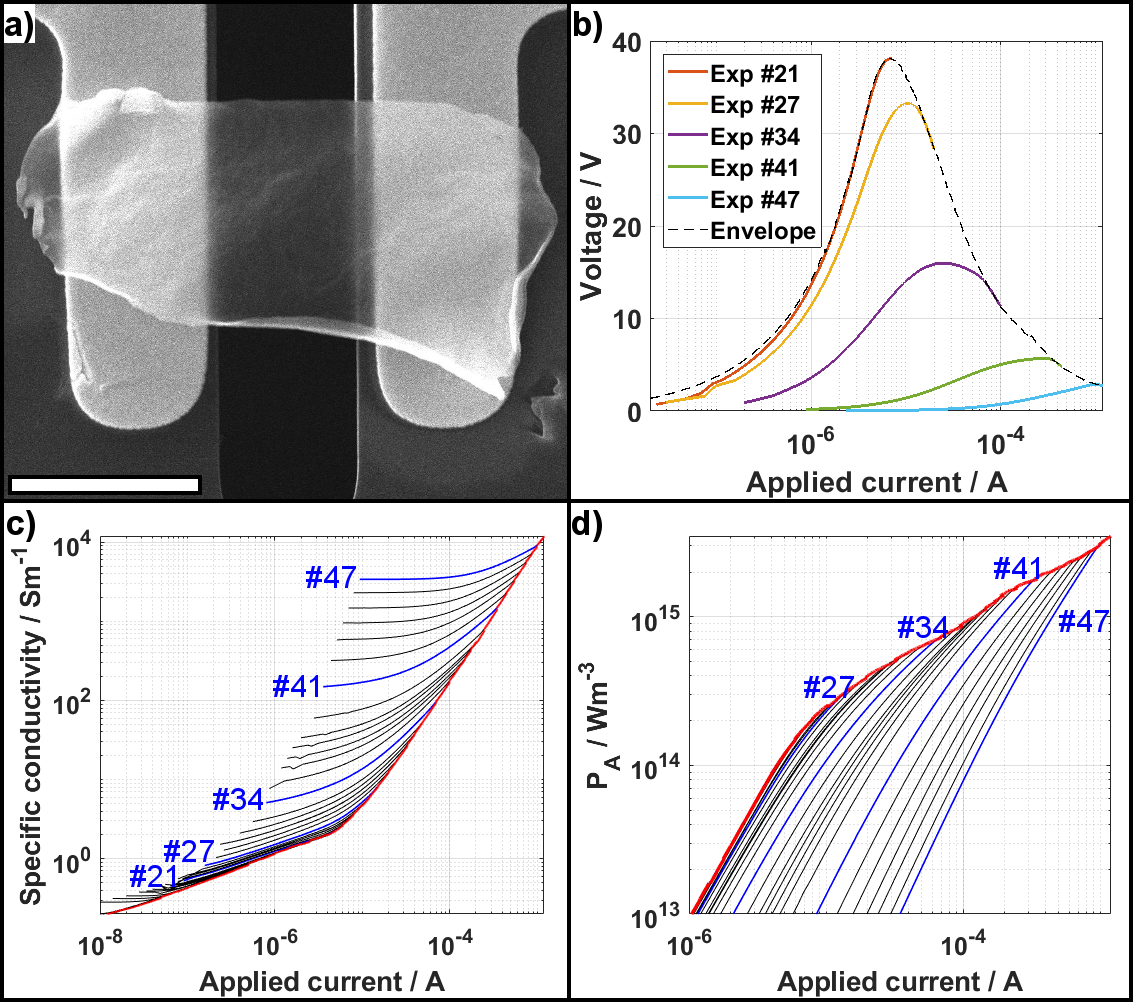}
\caption{(a) Scanning electron microscopy image of GO film contacted to the in-situ chip. Scale bar is 5~$\upmu$m. (b) Voltage over current curves displayed for selected measurements (solid lines) and an envelope containing the current ranges of each measurement exceeding the previous one (black dashed line) (c) Double-logarithmic plot of determined specific conductivity over applied current for all measurements (black) and envelope (red). Selected measurements are plotted and marked in blue. (d) Double-logarithmic plot of applied power density during the measurements (black) over the applied current with envelope (red) and selected measurements (blue).}
\label{F:Elec}
\end{figure}

\begin{figure}[ht]
\centering
\includegraphics[width=1\linewidth,keepaspectratio]{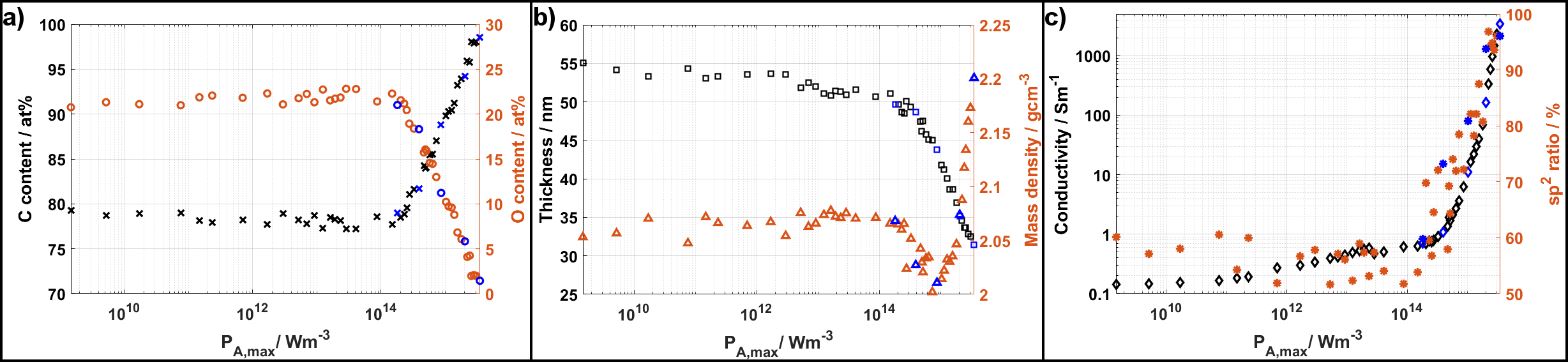}
\caption{Evolution of sample properties over maximum deposited power $P_{A,max}$. (a) C (black crosses) and O (red circles) content of sample, (b) thickness (black squares) and mass density (red triangles), (c) conductivity $\sigma_0$ (black diamonds) and sp\textsuperscript{2} ratio (red asterisks). Measurements \#21, 27, 34, 41, 47 are displayed with blue symbols.}
\label{F:Resu}
\end{figure}

Figure~\ref{F:Elec} shows the evolution of the electrical measurements. Five selected plots of the measured voltage over the applied current at logarithmic scale to better visualize the whole measurement range are displayed in Figure~\ref{F:Elec}b together with an envelope plot (dashed black line) that contains the parts of each measurement that exceed the previous $I_{max}$. The envelope is the approximation of performing all the measurements in a single run, i.e., applying a single linear current sweep. The initial V-I curves up to an applied current of 7~$\upmu$A (red curve in Fig.~\ref{F:Elec}b) overlap to a high degree and are displayed in linear scale as I-V curves in the SI for better comparison with literature data (Fig.~S2). Up to a voltage of 30~V, the current increases approximately with V\textsuperscript{2} before merging in an exponential increase and turning into a negative slope, thus a decreasing voltage with increasing current. Starting from this current range (above 7~$\upmu$A), the measured V-I curves start to deviate strongly from the envelope curve. The following measurements exhibit the same shape of the curve but with the maximum voltage decreasing and shifting to higher applied currents, implying a strong induced irreversible change in the GO film. Once the curves reach the envelope, that is the applied current is higher than $I_{max}$ of the previous measurement, the voltage starts to drop at an increased negative slope. The same trend is observed in the double-logarithmic plot of the film specific conductivity (Fig.~\ref{F:Elec}c)
calculated using the film size and its determined thickness (see Fig.~\ref{F:Resu}b). The measurements are plotted as black lines and the envelope (red curve) again contains the measurement parts that exceed the previous $I_{max}$. While only small changes are visible up to an applied current of 7~$\upmu$A that may be attributed to the higher applied current rather than to a modification of the material, a stronger increase in conductivity is observed starting approximately from measurement \#21 (7~$\upmu$A). In following measurements, the specific conductivity extrapolated to zero applied current $\sigma_0$, obtained by a linear fit to the initial values of each measurement, starts to increase noticeably. Throughout the whole experiment, the conductivity $\sigma_0$ increases more than four orders of magnitude from 0.12 to 3400~Sm\textsuperscript{-1}, thus changing from the insulating to the conductive regime. Although the measurements are conducted using a 2-probe setup, the influence of contact resistance (well below 50~$\Omega$) on the overall resistance (2~k$\Omega$ - 100~M$\Omega$) may be neglected.

GO and rGO may be seen as a network of conductive sp\textsuperscript{2}-hybridized islands separated by insulating barriers composed of oxidized and sp\textsuperscript{3}-hybridized or disordered sp\textsuperscript{2}-hybridized regions \cite{LocalChemicalStructure.2010,LocalCondAFM.2019}. Electrical conductivity and electron transport in GO and rGO is described by variable-range hopping (VRH) of electrons between the conductive islands based on tunneling processes \cite{GomezNavarro.2007,Kaiser.2009}. An increase in (bulk) GO conductivity when applying high voltages has been reported, which was assigned to a changing nature of the tunneling processes from direct to Fowler-Nordheim tunneling without inducing irreversible changes to the GO\cite{CondTunneling.2014}. The last measurements of our experiments confirm this increase in conductivity, which manifests itself in a change from a linear to an exponential increase in the I-V curves (Fig.~S2d).

For a meaningful presentation of the determined parameters and to be able to compare results between different experiment series, we calculated the applied power density $P_{A}$ during the measurements, which is displayed in Fig.~\ref{F:Elec}d and increases up to a maximum value of 3.5~10\textsuperscript{15}~Wm\textsuperscript{-3}. The plot is consistent with the V/I and conductivity evolution, as the curves first follow an almost identical line before an increasing current is needed to reach higher power values. The slope of the envelope curve is not continuous but increases between measurement \#36 and \#40 as well as after measurement \#47. 

\begin{figure}[ht]
\centering
\includegraphics[width=1\linewidth,keepaspectratio]{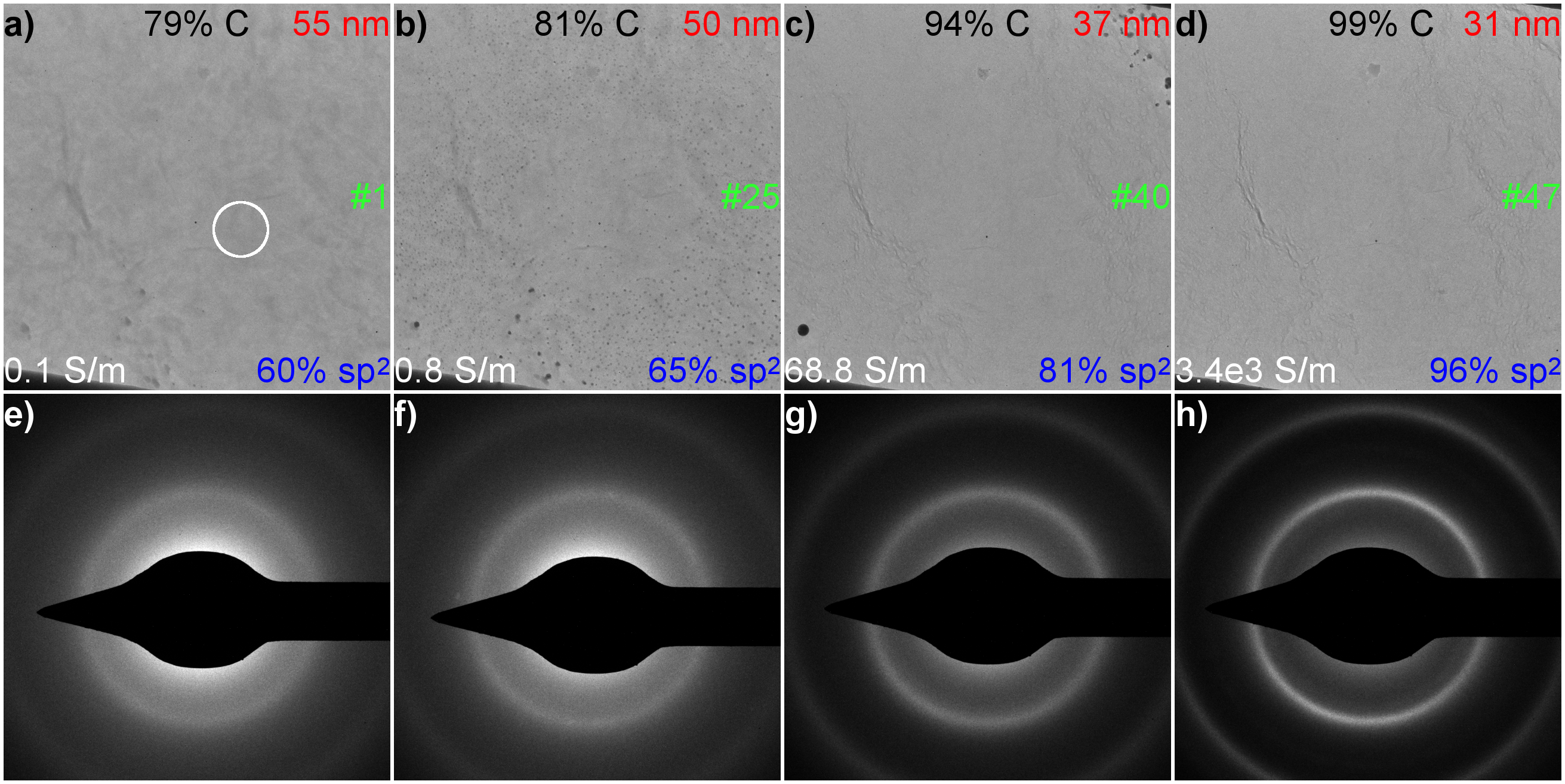}
\caption{Evolution of (a-d) TEM image and (e-h) corresponding diffraction pattern appearance at 4 measurement points. Corresponding values of C content (between 79 and 99\%), sample thickness (55 to 31~nm), measurement number, electrical conductivity (0.1 to 3.4~10\textsuperscript{3}~Sm\textsuperscript{-1}) and sp\textsuperscript{2} content (60 to 96\%) are displayed in the TEM images. The white circle in (a) marks the area investigated by EELS and ED. Size of TEM images is 3.5x3.5~$\upmu$m\textsuperscript{2} and of diffraction patterns 17x17~nm\textsuperscript{-2}. }
\label{F:TEM}
\end{figure}

Figure~\ref{F:TEM} shows four TEM images and diffraction patterns at specific measurement points. The film first appears amorphous and offers only marginal contrast that may arise from bending or local topographic changes, which are maintained throughout the whole experiment implying a constant film morphology. These morphological features of the film sharpen throughout the experiment.  After more than 20 measurements, small nanoparticles (NPs) start to form (Fig.~\ref{F:TEM}b). The NPs, supposedly containing Ga introduced during the FIB sample preparation and diffused on the sample surface, are present between measurement \#26 and \#40 but do not affect the determined flake properties (see section~SI3 of the SI). 

The evolution of the corresponding diffraction patterns is displayed in Figs.~\ref{F:TEM}e-h. The initial pattern reveals the presence of the [100] and [110] rings of the hexagonal graphitic lattice blurred by a smoothly decaying background stemming from amorphous contamination deposits introduced during the sample preparation process (Fig.~S3). After the FIB transfer, the slight texturing present in the initial drop-casted GO film is not visible anymore (Fig.~S3). The diffraction rings get sharper as the applied power density increases and appear highly homogeneous at the end of the experiment indicating the presence of a uniform film.

\begin{figure}[ht]
\centering
\includegraphics[width=1\linewidth,keepaspectratio]{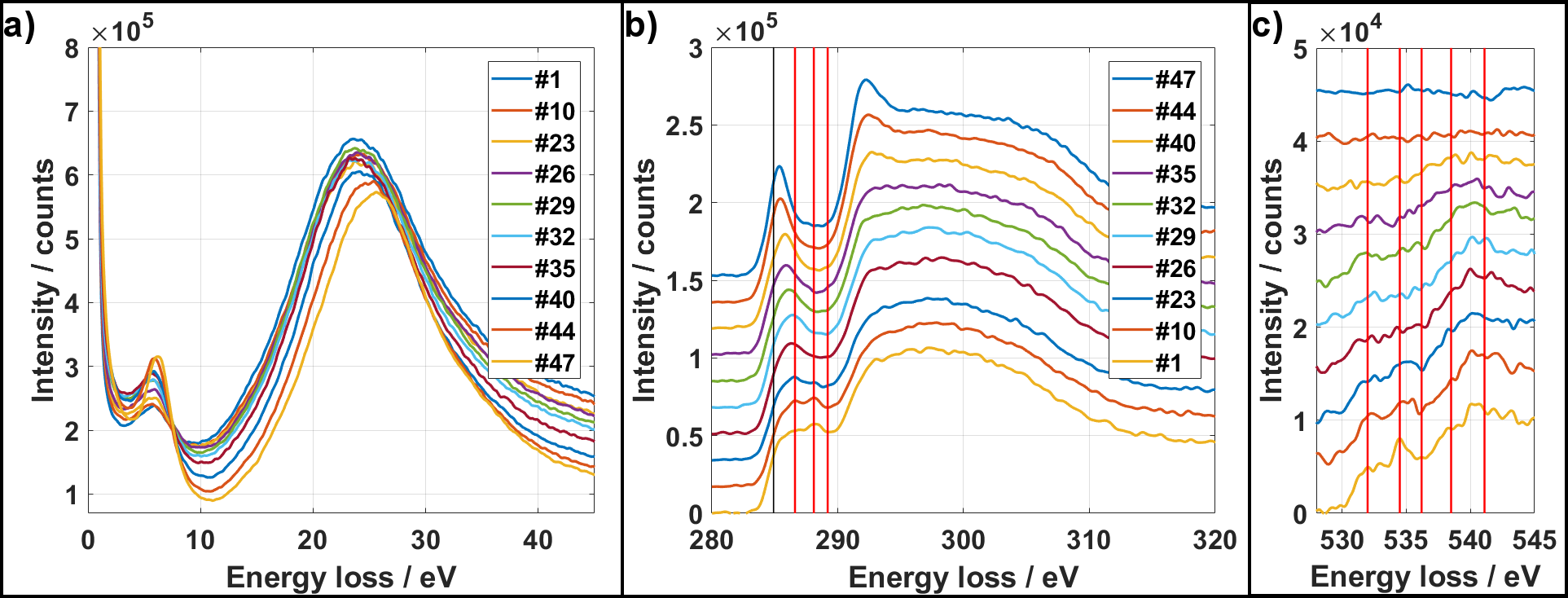}
\caption{Spectrum evolution of (a) the LL region revealing the surface and bulk plasmon, (b) the C-K edge and (c) the O-K edge with different contributions marked by vertical lines.}
\label{F:EELS}
\end{figure}

Figure~\ref{F:EELS} shows a comparison of EEL spectra at selected measurement points including the low-loss area (\ref{F:EELS}a), the C-K edge (\ref{F:EELS}b) and the O-K edge (\ref{F:EELS}c). The spectra have been normalized with the total beam current determined from the intensity of the ZL and LL region. A comparison of the whole acquired spectral range of the first and last measurement may be found in Figure~S4. 

The low-loss region (Fig.~\ref{F:EELS}a) exhibits two plasmonic features, i.e. the surface and the bulk plasmon of the sample, located between 6.5-7.5 eV and 26-27 eV, respectively. A sharpening of both plasmonic excitations is observed throughout the experiment. The surface plasmon shows a splitting in intermediate measurements and one of these surface plasmonic features may be attributed to the (Ga)NP presence. An overall decrease of scattered intensity in the low-loss region is observed, which may be clearly seen in the region between 40 and 50 eV. In addition, the bulk plasmon peak shifts to higher energies after measurement \#40.

The C-K edge in Figure~\ref{F:EELS}b shows a strong increase of the initial peak at 285.5~eV throughout the experiment while the following valley between 286-289~eV shows a decreasing intensity. The C-K edge is fitted using several functions assigned to different C-C ($\uppi$* and $\upsigma$*) and C-O bonds as detailed in the SI (Fig.~S5) \cite{Sp2MethodEELS.2008,EELSGO.2016,OxFuncEELS.2017,MarioGO}. The fit includes the $\uppi$* peak located at 284.9~eV (black vertical line in Fig.~\ref{F:EELS}b) assigned to the sp\textsuperscript{2}-hybridized C-C bonds, which is used for energy calibration. Note that, due to its shape, this peak exhibits its maximum at a higher energy \cite{Sp2MethodEELS.2008}. Different bonding type energies of C with O are marked with red vertical lines and correspond to the positions of the different functions employed in the fitting procedure. An additional peak at 291.5~eV develops on top of the broad excitation in the last measurements, typical for sp\textsuperscript{2}-rich C \cite{Sp2MethodEELS.2008,Zhang.2016}. The O-K edge (Figure~\ref{F:EELS}c) shows a gradual decreasing overall intensity starting from $P_A$=3$\cdot$10\textsuperscript{14}~Wm\textsuperscript{-3} (\#26), which vanishes completely in the final spectrum. The five vertical lines positioned at 532, 534.5, 536.2, 538.5 and 541.1~eV separate the edge into four regions. The evolution of the different peak intensities of C-K and O-K are displayed in the SI (Figure~S6) and is discussed later jointly with their assignation to functional groups. 

The EELS data is analyzed as detailed in the SI (section SI6) to obtain an estimation of the thickness via the inelastic mean free path \cite{Egerton.2011}, an estimation of the mass density derived from the bulk plasmon energy \cite{FerrariMassDensity.2000,LajaunieEELSaCH.2017,DLC_nearFrictionlessLiu.2007} and to quantify the film composition (O and C)   \cite{Egerton.2011}. The sp\textsuperscript{2} content is estimated by determining the intensity ratio of the $\uppi$*-peak intensity $I_{\pi*}$ at 284.9~eV and the total C-K edge intensity integrated over the range between 282 and 320~eV and comparing the ratio with a HOPG reference sample in [001] orientation acquired using the identical setup (details in the the SI, section~S6).

Figure~\ref{F:Resu} shows the plots of six parameters over $P_{A,max}$, including the complementary contents of C and O (Fig.~\ref{F:Resu}a), the thickness and mass density (\ref{F:Resu}b) and the specific conductivity $\sigma_0$ together with the sp\textsuperscript{2} ratio (\ref{F:Resu}c). The plots may be separated in two phases: The first phase goes up to 2$\cdot$10\textsuperscript{14}~Wm\textsuperscript{-3} and covers the first 20~measurements in which only slight changes may be observed. The C content decreases while the O content increases by approximately 1\%, the thickness goes down from 55 to 50~nm, while the mass density slightly increases. The conductivity shows a continuous increase by approximately an order of magnitude whereas the sp\textsuperscript{2} ratio remains in the same range between 50 and 60\%. 

The second phase starts from 2$\cdot$10\textsuperscript{14}~Wm\textsuperscript{-3} (\#21) and is characterized by a continuous increase/decrease of the C and O content, respectively. The thickness starts to decrease at a higher slope from 4$\cdot$10\textsuperscript{14}~Wm\textsuperscript{-3} (\#27) until reaching a final thickness of 31.5~nm, which signifies a decrease of more than 40\% with respect to the initial thickness. The mass density however offers an additional phase as it first decays to 2~gcm\textsuperscript{-3} in measurement \#33 corresponding to  $P_{A,max}$~=~7$\cdot$10\textsuperscript{14}~Wm\textsuperscript{-3} before it rises reaching a value of 2.2~gcm\textsuperscript{-3} as final value. This value is similar to the mass density of graphite \cite{GraphiteDelhaes.2014}. The conductivity and the sp\textsuperscript{2} ratio both follow a strong and continuous rise in the second half of the experiment reaching values of 3.4~$\cdot$~10\textsuperscript{3}~Sm\textsuperscript{-1} and 96\% sp\textsuperscript{2} hybridization. 

The experimental results clearly show that we successfully performed the in-situ reduction of a GO film by Joule heating as all the monitored parameters show a clear trend towards rGO properties, that is, a higer conductivity and sp\textsuperscript{2} ratio and a decreasing O content. Moreover, a comparison of EEL spectra acquired at different positions of the finally reduced film reveals that the whole GO film was reduced to the same degree (SI, Figure~S7). By contacting a rectangular GO film at various points over its width to the chip electrode, an uniform distribution of the current density within the film was achieved (Fig.~S1). In contrast, we observed different degrees of reduction within a similar film where a more point-like contacting of the film or a non-symmetric geometry of the film was used, which may be attributed to a variation of the current density within the GO film. We performed an additional XPS study of the GO before and after the experiment, which also proves the successful reduction (Fig.~S8). In addition to the single experiment described in the main manuscript, Figure~S9 shows the comparison of the results obtained from the in-situ reduction and conductivity measurements of three different films. The overall evolution is in very well agreement and the onset power density from which the GO reduction is induced depends on the initial degree of oxidation of the films.

The experiment can be divided in two phases: before and after reaching the necessary power density of 2$\cdot$10\textsuperscript{14}~Wm\textsuperscript{-3} to initiate a strong irreversible reduction process of the sample. The first phase is characterized by a slight increase in relative O content and mass density, a decreasing thickness and a slightly improved conductivity. In addition, a redistribution of the intensity in the O-K and C-K EEL edges may be observed.  While during this phase the power deposited in the sample is not yet high enough to break stronger bonds and induce a reduction, it is sufficient to promote several mechanisms, which are proposed to be responsible for the observed changes in the material.

TEM samples always carry several adsorbed materials. While water molecules are typically adsorbed to the sample surface with higher bonding energies (chemisorption), hydrocarbon contamination is only weakly bound to the sample surface via physisorption \cite{Hettler.2018}. In the present experiment, the film may be seen as a core of a stack of several GO flakes surrounded by a carbonaceous contamination layer deposited during the FE/FIBID process by dissociation of adsorbed precursor molecules by secondary electrons generated during the contacting process and by the minimum amount of primary electrons necessary for imaging of the sample. Water molecules are thus abundant, both intercalated in the GO \cite{IntercalWater.2010,GPOCNTsJ.D.Nunez.2017} as well as chemisorbed on the deposited contamination layer. The electrical current together with the electron irradiation of the sample induce two effects in the film: weakly bond hydrocarbon molecules might be stimulated to desorb leading to the increasing O content of the sample. Water molecules, chemisorbed on the surface or intercalated within the GO sheets, are too tightly bond to the sample to be forced to desorb but might however diffuse within the sample driven by the applied electrical fields. 

Considering these aspects, we assign the peaks at 288.1 and 534.5~eV, strongly visible in the initial C-K and O-K spectra, to adsorbed or intercalated water molecules. Due to diffusion, the water molecules start to bind more strongly to the carbon matrix at suitable sites resulting in the broadening of the peaks in the first phase. The assignation of the peak to intercalated/adsorbed water is supported by the analysis of the rGO film after rehydration at air (Fig.~S10), where the O-K edge as well shows a peak at 534.5~eV. In the C-K edge, the intensity moves from the peak at 288.1~eV to the C-O/H peak at 286.6~eV indicating a stronger bond of the intercalated water to the C matrix. Again, a look on the rehydration data of the C-K edge confirms the assignation of adsorbed/intercalated water to the peak at 288.1~eV, as this peak gains importance during the rehydration process (Fig.~S10). In summary, the diffusion-induced restructuring seems to go in hand with a minor densification of the film, leading, together with the desorption of hydrocarbon molecules, to the increased mass density, a smaller thickness of the sample and the increased conductivity. Additional studies, such as an EELS analysis during the application of electrical power, are necessary to shed light on the nature of the processes evolved in the first phase.

Reduction via Joule heating has the advantage that the energy is deposited in a more localized manner within the film. As GO is well described by a sp\textsuperscript{2} network distorted by resistive islands of OFGs and/or sp\textsuperscript{3} hybridized regions, the film resistance is given by a sum of many small resistances through which the conducting electrons have to tunnel \cite{CondTunneling.2014}. During the first phase the local temperature increase is not high enough to cause an actual reduction and intercalated water molecules are only caused to diffuse to a less resistive site within the GO film instead of being caused to evaporate once reaching temperatures around 100\degree C in thermal annealing experiments \cite{InterCalRouziere.2017}.

The second phase, starting from an applied power density of 2$\cdot$10\textsuperscript{14}~Wm\textsuperscript{-3}, is characterized by strong changes in all observed parameters. Now, the power applied to the sample is high enough to locally break bonds and reduce the GO. The O content and thickness decrease while sp\textsuperscript{2} ratio and conductivity increase. Only the mass density first shows a drop before strongly rising. An explanation may be that O atoms are removed before the thickness of the film starts decreasing, which leads to a decreasing mass density. Only after a stronger reduction, the distance between sp\textsuperscript{2} hybridized sheets starts to decrease allowing a densification of the sample. In fact, the O content already starts to decrease strongly from 2$\cdot$10\textsuperscript{14}~Wm\textsuperscript{-3} (\#21), while a stronger thickness decrease is observed from 4$\cdot$10\textsuperscript{14}~Wm\textsuperscript{-3} (\#27). Assuming a distance of 0.33 nm between the C sheets in the final reduced film, the number of layers can be estimated to 95. Assuming the same amount of layers in the initial film - we did not observe a significant loss of C during the experiment  - the interplanar distance of the GO film amounts to 0.55-0.6 nm, which is in agreement with values typically encountered in GO \cite{InterCalRouziere.2017}. As the amount of C remains largely constant throughout the experiment, we assume that the amorphous C deposits are graphitized and taken up in the rGO film, similar to the process previously reported for graphene samples \cite{Barreiro.2013}.

A closer look at the electron energy-loss near edge structure (ELNES) of the C-K and O-K edges shows that the peaks previously assigned to adsorbed/intercalated water bond to the C matrix (288~eV and 534~eV) are the first ones to decrease in phase two (see also Figure~S6). This is in agreement with literature, as water is the first species removed during GO reduction \cite{IntercalWater.2010,GPOCNTsJ.D.Nunez.2017}. Previous studies on the ELNES of the C-K edge and the O-K edge for different types and amounts of OFGs do not report this assignation as they were conducted on dried GO samples \cite{OxFuncEELS.2017,EELSGO.2016,MarioGO}. 

The peak at 532~eV is the next contribution that starts decreasing in the reduction process, which we assign to epoxide (C-O-C) groups as thermal reduction experiments show that these OFGs are the first to disappear already at intermediate temperatures \cite{AcikThermalReductionIR.2011,GangulyThermalRedXPS.2011} or at low applied voltages in Joule heating\cite{Xu.2011}. Indeed, DFT calculations indicate a major contribution of epoxides to the lower energy range of the O-K ELNES \cite{OxFuncEELS.2017}. The spectral range between 536 and 541~eV may then be assigned to hydroxyl and carboxyl groups (C-O-H, COOH), which partially resist thermal reduction up to 1000\degree C \cite{GangulyThermalRedXPS.2011}. The difference between peak energies C-O-H and C-O-C bonding types within the C-K edge is very small and is not resolved in this study. Both OFGs are summarized in the peak at 286.6~eV, which shows a continuos decrease throughout the reduction process before being almost absent in the final spectrum. The COOH peak at 289.2~eV reduces its intensity only in the last part of the experiment.

The order of the reduction of the OFG contributions throughout the experiments is very similar to GO reduction by thermal annealing \cite{AcikThermalReductionIR.2011,GangulyThermalRedXPS.2011}. Nevertheless, there exist considerable differences between thermal annealing and Joule heating. In the presented experiment, all monitored parameters in phase 2 show a smooth evolution. This smoothness is thought to be caused by the localized nature of the Joule heating process, which causes a spatial temperature distribution that might strongly vary depending on the local resistivity. As the resistivity of the individual barriers not only depends on its composition and the types of OFGs but also on its size and distribution, the energy necessary for the reduction of an individual OFG depends on its local environment. Therefore the reduction of an OFG occurs during a range of applied power densities leading to the smooth evolution of the monitored parameters. This is in contrast to GO reduction by thermal annealing, where the entire sample exhibits a single temperature and all OFGs are reduced simultaneously at a determined temperature causing a step-wise reduction process.

We performed calculations of the Joule heating process with COMSOL Multiphysics assuming a homogeneous resistivity to estimate an average overall temperature of the film. The simulations showed a strong dependence of the film temperature on the thermal conductivity (Fig.~S12), being between slightly above room temperature for values typical for graphene up to more than 1000\degree C for experimental values of GO \cite{ThermalCondTheoMu.2014,ThermalCondExpKim.2018}. Although we did not perform a direct temperature assessment during the measurements, the degree of reduction suggests (local) temperatures of $>$1000\degree C reached by the Joule heating process.  

In conclusion, the study successfully links the reduction state and the microscopic structure of GO films with its electrical conductivity. Within the in-situ TEM experiment, a GO film with an initial thickness of 55 nm, an O content of more than 20 at.\% and a specific conductivity of 0.15~Sm\textsuperscript{-1} is reduced by Joule heating in a controlled manner to a highly reduced GO with a thickness of 31 nm, an O content of $<$1 at.\% and a specific conductivity of 3.4~$\cdot$10\textsuperscript{3}~Sm\textsuperscript{-1}. The final film exhibits a mass density close to graphite and the sp\textsuperscript{2} content is increased strongly throughout the experiment. The analysis of the EELS spectra allows the assignation of a specific feature to intercalated and adsorbed water and the determination of the evolution of different OFGs throughout the reduction process. Further studies are required to elucidate a more detailed analysis of the involved processes that are enticed during the application of the electrical currents necessary for reduction by Joule heating.

\begin{acknowledgement}
S.H. and R.A. acknowledge funding by German Research Foundation (DFG project He 7675/1-1). Research supported by the Spanish MINECO and MICINN (MAT2016-79776-P, AEI/FEDER, EU; PID2019-10473P6B-100/AEI/10.13039/501100011033; ENE2016-79282-C5-1-R, AEI/FEDER, EU and
PID2019-104272RB-C51/AEI/10.13039/501100011033), Government of Aragon (projects DGA E13-17R (FEDER, EU); DGA T03-17R (FEDER, EU) and DGA T03-20R) and European Union H2020 programs “ESTEEM3” (823717), Flag-ERA GATES (JTC - PCI2018-093137), ITN Enabling Excellence (MSCA-ITN) and Graphene Flagship (881603). 
We thank L. Casado and G. Antorrea (LMA, U. Zaragoza, Spain) for help with the micro-needle assisted sample preparation and with XPS acquisition, respectively. All the characterization studies have been carried out at the LMA, U. Zaragoza, Spain.
\end{acknowledgement}

\begin{suppinfo}
\label{SI}
Supplementary information file contains additional experimental results.
\end{suppinfo}

\bibliography{biblio}

\end{document}